# Generation of optical toroidal vortex with circular asymmetric gratings

**WEICHAO LIU,**[1,2] **JIE CHENG,**[1,2] **CHENHAO WAN,**[1,*]

[1]*School of Optical and Electronic Information and Wuhan National Laboratory for Optoelectronics, Huazhong University of Science and Technology, Wuhan, Hubei 430074, China*
[2]*These authors contributed equally: Weichao Liu, Jie Cheng.*
*[wanchenhao@hust.edu.cn](wanchenhao@hust.edu.cn)*

**Toroidal vortex, a topological structure commonly observed in nature, exist in various types such as bubbles produced by dolphins and the air flow surrounding a flying dandelion. A toroidal vortex corresponds to a spatiotemporal wave packet in the shape of a donut that propagates in the direction perpendicular to the plane of the ring. In this work, we propose a circular asymmetric grating to generate vortex rings. A cylindrical vector wave packet is transformed by the device into a transmitted toroidal vortex pulse. Such a compact toroidal vortex generator may find applications in optical topology research and high-dimensional optical communications.**

The pioneering work of Allen and his colleagues[1] discovered that light with a spiral phase structure $e^{il\varphi}$ possess an orbital angular momentum (OAM) of $l\hbar$ for each photon, where $l$ is the topological charge, $\hbar$ is the Dirac constant, and $\varphi$ is the azimuthal angle. Due to this unique azimuthal phase structure, vortex beams with different topological charges are mutually orthogonal, introducing a new dimension $l$ which is unbounded in theory. Consequently, vortex beams have found a wide range of applications in high-capacity optical communication systems[2-4], optical tweezers[5, 6], optical holographic encryption[7-10] and high-dimensional quantum systems[11, 12].

Contrary to vortex beams that have a spiral phase in the cross section, spatiotemporal optical vortices (STOVs) feature a spiral phase structure situated in the spatiotemporal plane, thereby possessing transverse orbital angular momentum perpendicular to the propagation direction[13, 14]. STOVs can be generated with a two-dimensional pulse shaper or metasurface and characterized by scanning or single-shot methods[15-21].

The toroidal vortex can be considered as a high-dimensional STOV. Slicing in any radial direction will always obtain a cross section $(r-r_0,t)$ that represents a STOV. A toroidal vortex can be generated through wrapping an elongated STOV into a ring shape using optical conformal mapping[22].

In this work, we propose a scheme based on a circular and asymmetric grating structure for the generation of optical toroidal vortex in a compact way. Numerical simulations demonstrate that this structure is capable of transforming a normal incident cylindrical vector wave packet into a toroidal vortex. The device is designed for the C-band, holding potentials for applications in high-dimensional optical communications as well as quantum key distributions.

The toroidal vortex of light, a three-dimensional spatiotemporal structure, can be represented in the reference frame as:

$$E(x,y,\tau) = \left( \frac{\sqrt{(r-r_0)^2 + \tau^2}}{w_0} \right) \exp\left( -\frac{(r-r_0)^2 + \tau^2}{w_0} \right) \exp\left( -il\tan^{-1}\left( \frac{\tau}{r-r_0} \right) \right), \quad (1)$$

where $(x,y)$ represents the Cartesian coordinates of the spatial plane, $r=\sqrt{x^2+y^2}$ is the radius of the spatial plane, $\tau$ is the normalized retarded time, $w_0$ is a constant representing the size of the vortex ring in the spatiotemporal plane.

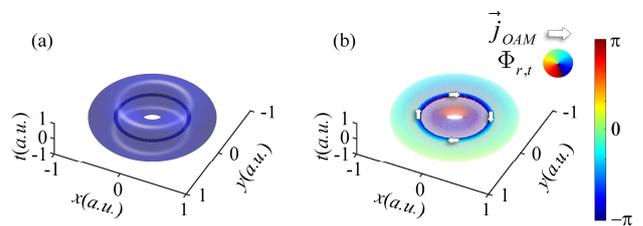

Fig. 1. (a) Iso-intensity and (b) phase distribution of a toroidal vortex. The arrow indicates the local OAM density ($\vec{j}_{OAM}$) in Fig. 1(b). The colorbar indicates the phase range in Fig. 1(b).

The iso-intensity surface and phase distribution of a toroidal vortex is shown in Fig. 1. Due to the spiral phase in the spatiotemporal plane ($r$-$r_0$, $\tau$), the direction of local OAM density

$\vec{j}_{OAM}$ is always perpendicular to the propagation direction and circulates in the cross-section plane. The phase singularities form a circular singularity line at $r = r_0$. The production of a vortex ring relies on the custom generation of this circular phase singularity line in the spatiotemporal domain.

Breaking the mirror symmetry is the key to introducing a phase singularity in the spatiotemporal plane[16]. Without loss of generality, one considers a one-dimensional periodic grating structure with the plane $x = 0$ serving as the mirror-symmetric mirror. For a normal incident pulse with Gaussian distributions in the $(x,t)$ plane, the phase distribution of the transmitted pulse must also be symmetric about the $x = 0$ plane due to the symmetry of the grating structure, i.e., phase singularities cannot be generated. Therefore, in order to generate STOVs, the mirror symmetry about $x = 0$ of the structure must be broken. The proposed structure should not have mirror symmetry in any direction.

The toroidal vortex can be seen as a log-polar to Cartesian transformation from an elongated STOV. To generate a toroidal vortex, a one-dimensional asymmetric periodic grating can be extended in the direction perpendicular to the grating vector and wraps around into a circular shape. As shown in Fig. 2, nano rings with different radii are densely arranged into a disk. In any radial direction, the structure is a one-dimensional asymmetric periodic grating. This device is composed of a material with a relative permittivity of 12 at 1550 nm, a value similar to silicon's relative permittivity in this wavelength range. The distance between the inner and outer radii of each ring is $dr = 1291.2$ nm. $N$ rings are assembled to form a large circular disc. If the initial ring radius is $r_0$, then the inner radius of each ring is $r_n = r_0 + (n-1)dr$ where $n = 1, 2, 3, \cdots, N$. In each ring, to break spatial mirror symmetry, three concentric grooves are etched on a ring-shaped base with a width of $dr$, creating two non-identical concentric ridges with widths $w_1 = 206.6$ nm, $w_2 = 557.8$ nm, and thicknesses $h_1 = 501$ nm, $h_2 = 206.6$ nm. The spacing between the two concentric ridges is $g = 82.6$ nm. The thickness of the substrate is $t = 100$ nm.

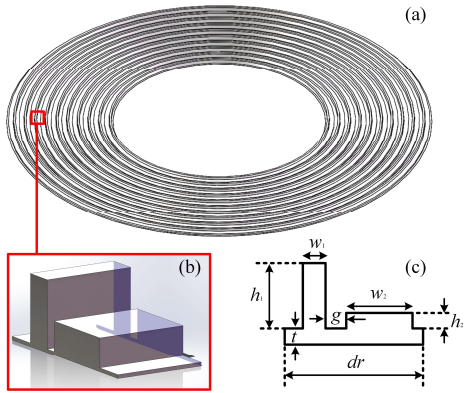

Fig. 2. Device Schematic: (a) Overall schematic diagram of the device. (b) Schematic diagram of the device unit. (c) Cross-sectional schematic diagram of the device, with dimensions annotated.

For a normal incident, azimuthally polarized, cylindrical vector wave packet, the electric field can be expressed as:

$$\vec{E}_{in} = \begin{bmatrix} \cos\left(\varphi + \dfrac{\pi}{2}\right) \\ \sin\left(\varphi + \dfrac{\pi}{2}\right) \end{bmatrix} A_{in}(r,t) e^{-i\omega_0 t}, \quad (2)$$

where $\begin{bmatrix} \ \end{bmatrix}$ is the Jones matrix, $\begin{bmatrix} 1 \\ 0 \end{bmatrix}$ represents the electric field along the $x$ direction, $\begin{bmatrix} 0 \\ 1 \end{bmatrix}$ is the electric field along the $y$ direction, $\varphi = \tan^{-1}\left(\dfrac{y}{x}\right)$ is the azimuthal angle in the $(x,y)$ plane, $\omega_0 = \dfrac{2\pi c}{\lambda_0}$ is the center frequency, $\lambda_0 = 1550$ nm, and $A_{in}(r,t)$ is the envelope of the incident electric field.

We decompose the envelope of the input pulse into a series of plane waves via Fourier transformation:

$$A_{in}(r,t) = \iint \tilde{A}_{in}(k_r, \Omega) \exp(ik_r r - i\Omega t) dk_r d\Omega, \quad (3)$$

where $\tilde{A}_{in}(k_r, \Omega)$ is the angular spectrum of envelope, $\Omega = \omega - \omega_0$ is the sideband angular frequency with respect to the center frequency $\omega_0$, and $k_r$ is the radial wavevector. The transmission spectrum function is determined by $H(k_r, \Omega) \equiv \dfrac{\tilde{A}_{out}(k_r, \Omega)}{\tilde{A}_{in}(k_r, \Omega)}$, where $\tilde{A}_{out}(k_r, \Omega)$ is the angular spectrum of output envelope. Since $\dfrac{2\pi}{dr} > \dfrac{2\pi}{\lambda_0}$, $\lambda_0$ is the wavelength in vacuum, the circular asymmetric grating generates only zero-order diffracted pulse. Therefore, the transmission spectrum function of zero-order diffracted pulse is derived.

Fig. 3(a) and 3(b) show the amplitude and phase distributions of the transmission spectrum function with respect to radial wavevector $k_r/k_0$ and local frequency $\Omega/\omega_0$ with the help of COMSOL. It can be seen that the amplitude of the transmission spectrum function $Abs(H(k_r, \Omega))$ exhibits a good linear relationship with $k_r$ and $\Omega$ around $-0.15 \leq k_r/k_0 \leq 0.15$ and $-0.01 \leq \Omega/\omega_0 \leq 0.01$. The transmission spectrum function is assumed to take the following form:

$$H(k_r, \Omega) = C_r k_r + C_t \Omega, \quad (4)$$

where $C_r$ and $C_t$ are two constants. Then, the output pulse reads:

$$\begin{aligned} A_{out}(r,t) &= FT^{-1}\{\tilde{A}_{in}(k_r,\Omega) H(k_r,\Omega)\} \\ &= \left(-iC_r \dfrac{\partial A_{in}(r,t)}{\partial r} + iC_t \dfrac{\partial A_{in}(r,t)}{\partial t}\right) \end{aligned} \quad (5)$$

where $FT^{-1}\{\}$ represents the inverse Fourier transform. This indicates that the output pulse will be the first-order differentiation of the input pulse in spatial and temporal domains.

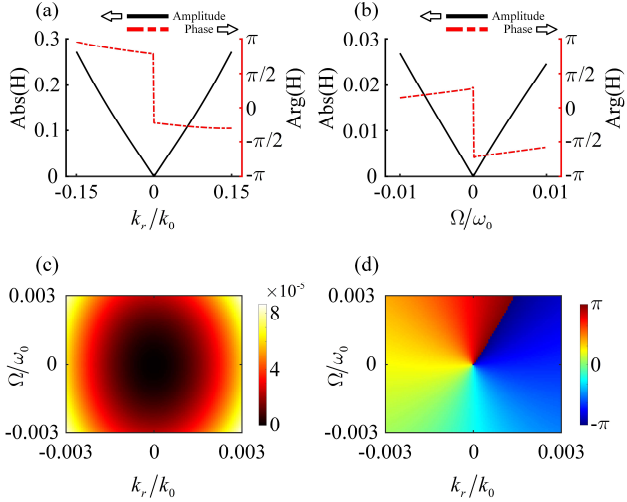

Fig. 3. Transmission spectrum function of circular asymmetric grating. The amplitudes and phase function along (a) $\Omega = 0$ and (b) $k_r = 0$; (c) Amplitude and (d) phase distribution of the transmission spectrum function with respect to $k_r$ and $\Omega$.

From Fig. 3(d), it can be seen that the phase of the transmission spectrum function exhibits a typical spiral phase form with a topological charge equal to -1 and the amplitude varies only radially. Therefore, the angular spectrum of the output envelope can be expressed as:

$$\widetilde{A}_{out}(k_r,\Omega) = g(r')\widetilde{A}_{in}(k_r,\Omega)\exp(-i\theta), \qquad (6)$$

where $(r',\theta)$ are the polar coordinates in the plane $(k_r/k_0, \Omega/\omega_0)$, $r' = \sqrt{\left(\frac{\Omega}{\omega_0}\right)^2 + \left(\frac{k_r}{k_0}\right)^2}$ and $\theta = \tan^{-1}\left(\frac{\Omega}{ck_r}\right)$, $abs()$ denotes the absolute value. For a normal incident cylindrical vector Gaussian pulse, the envelope distribution in the $(r,\tau)$ plane is independent of the azimuthal angle, thus $\widetilde{A}_{in}(k_r,\Omega) = \widetilde{A}_{in}(r')$. After the Fourier transformation, the envelope of the output pulse can be written as[23]:

$$\begin{aligned}A_{out}(\rho,\phi) &= FT^{-1}\{g(r')\widetilde{A}_{in}(r')\exp(-i\theta)\}\\ &= -i2\pi \times \exp(-i\phi) \times H_l\{g(r')\widetilde{A}_{in}(r')\}\end{aligned}, \qquad (7)$$

where $H_l\{g(r')\widetilde{A}_{in}(r')\} = \int_0^\infty r'g(r')\widetilde{A}_{in}(r')J_1(2\pi\rho r')dr'$, $J_1()$ is the first kind Bessel function, $\rho = \sqrt{r^2 + \tau^2}$ and $\phi = \tan^{-1}\left(\frac{\tau}{r-r_0}\right)$ are the polar coordinates in spatiotemporal plane $(r,\tau)$. From Eq.(7), one can observe that the transmitted pulse possess a spiral phase in the spatiotemporal plane, with phase singularities located at $r = r_0$, i.e., a distinctive toroidal topological structure is generated.

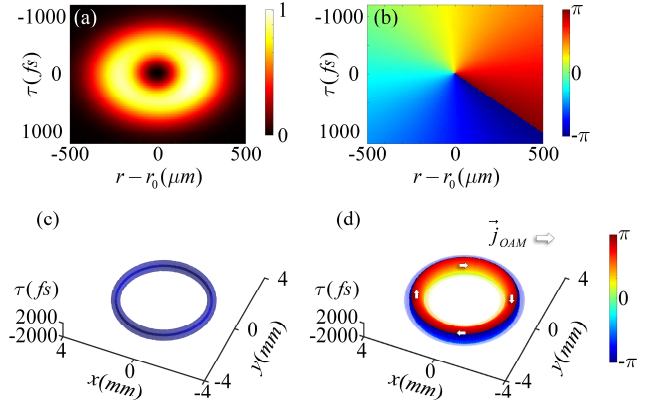

Fig. 4. (a) Amplitudes and (b) Phase distribution of the output pulse in spatiotemporal domain $(r,\tau)$. (c) Iso-intensity and (d) phase distribution of in three-dimensional space $(x,y,\tau)$. The arrow indicates the local OAM density ($\vec{j}_{OAM}$) in Fig. 4(d). The colorbar indicates the intensity range and phase range respectively.

As shown in Fig. 4(a) and 4(b), the amplitude and phase distributions of the output pulse in spatiotemporal plane $(r,\tau)$ are depicted. It can be observed that in the spatiotemporal plane $(r,\tau)$, the intensity distribution of the output pulse exhibits a hollow "doughnut" shape, while the phase varies uniformly along the azimuthal direction $\phi$. These are typical characteristics of STOVs. In three-dimensional space $(x,y,\tau)$, it can be seen in Fig. 4(a) and 4(b) that the output pulse exhibits a zero-intensity point at $r = r_0$, and the direction of local OAM density $\vec{j}_{OAM}$ is perpendicular to the propagation direction and rotates along the circular ring $r = r_0$, displaying the unique features of a toroidal vortex.

Finally, we analyzed the feasibility of fabricating the toroidal vortex generator. Since it is impractical to fabricate perfect circular ring devices, in practice, each unit will be rotated by a certain angle and connected to form a circular ring, as shown in Fig. 5(a) and 5(b). To address this, we analogize the concept of duty factor in one-dimensional grating devices and propose an equivalent duty factor parameter $\eta$ to quantify the difference between the practical device and the ideal circular ring. The parameter $\eta$ can be expressed as:

$$\eta = \frac{S_2 - 2S_1}{S_2}, \qquad (8)$$

where $S_1$ is the area of a single unit in the $(x,y)$ plane, and $S_2$ is the area corresponding to the ideal sector region after connecting two units.

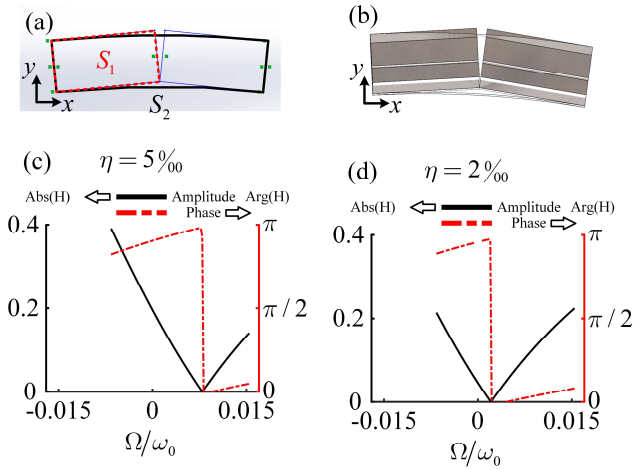

Fig. 5. The (a) plan view and (b) three-dimensional view of two units fitting the sector region. The amplitudes and phase function of transmission spectrum function along $k_r = 0$ for equivalent duty factor $\eta$ equal to (c) $5‰$ and (d) $2‰$.

As shown in Fig. 5(c) and 5(d), at $\eta = 5‰$ and $\eta = 2‰$, both the lowest points of the amplitude and the discontinuity points of the phase in the transmission spectrum function will shift. This implies that the central wavelength of the toroidal vortex generator will change. However, when $r_0 > 5dr$, the parameter $\eta < 2‰$. At this point, the change in the central wavelength is rather small. As to the device we designed, $r_0 > 200dr$, thus the change in the central wavelength caused by fabrication is negligible.

In conclusion, we propose a circular asymmetric grating to generate the toroidal vortex of light at 1550 nm. We utilized COMSOL to calculate the transmission spectrum function and obtained the intensity and phase distribution of the transmitted pulse. For a normal incident cylindrical vector Gaussian wave packet, this device is capable of producing a toroidal vortex of light with phase singularities located at $r = r_0$. The feasibility of fabrication is analyzed. We believe this device provides a platform for studying spatiotemporal topology in optics and may find applications in high-dimensional optical communications.

**Funding.** National Natural Science Foundation of China (62375099, 62335009).

**Disclosures.** The authors declare no conflicts of interest.

**Data availability**

Data underlying the results presented in this paper are not publicly available at this time but may be obtained from the authors upon reasonable request.